\documentclass[11pt, a4paper]{article}
\usepackage{graphicx}
\usepackage{amssymb}
\usepackage{stmaryrd}
\usepackage{enumitem}

\usepackage{geometry}
\usepackage{tikz}
\usepackage{float}
\usepackage{url}
\usepackage{mathtools}
\usepackage{amsmath,amsfonts,amsthm}
\usepackage{mathrsfs}

\usepackage{pgfplots}
\pgfplotsset{compat=1.18}
\usepackage{stmaryrd}
\usepackage{booktabs}
\usepackage{array}
\usepackage{caption}
\usepackage{xcolor}
\usepackage{multirow}
\usepackage{siunitx}

\usepackage{subfig}

\usepackage{ytableau}
\usepackage[colorlinks,
linkcolor=blue,
anchorcolor=blue,
citecolor=blue
]{hyperref}
\title{Integrating Linear Regression and Multi-Criteria Decision Making for Assessing Financial Statement Risks in Manufacturing Firms}
\author{
Duaa Abdullah$^{a}$ \thanks{Corresponding author: duaa1992abdullah@gmail.com
}  \, \& \, Marwa Abdullah$^{b}$ \\ $^{a}$ Physics and Technology School of Applied Mathematics and Informatics \\
Moscow Institute of Physics and Technology, 141701, Moscow region, Russia
\\
$^{b}$ Department of Finance, State University of Management, 99 Ryazansky Prospekt,\\ 109542, Moscow, Russia.
}

\date{}
\begin{document}
\maketitle
\begin{abstract}
Evaluating the financial performance of manufacturing firms requires consideration of both the time value of money and the relative importance of multiple decision criteria. Conventional approaches relying solely on deterministic discounting often fail to account for interactions among economic, operational, and managerial factors. This study proposes an integrated framework that combines time-discounted economic analysis with linear regression to evaluate control system efficiency.
A theoretical discounting model is first developed to convert costs and benefits occurring at different times into present-value terms using compound interest functions. The model accommodates one-time expenditures, time-proportional costs, and complex cost structures arising during system development and commissioning. To empirically assess how discounted economic performance is influenced by multiple criteria, linear regression serves as the approximation method.

Expert-derived Average Weight Scores (AWS) for secondary and primary evaluation criteria are used as explanatory variables in the regression model. These weights reflect the relative economic importance of factors such as cost structure, investment risk, reliability, and cash-flow characteristics. The regression results quantify each criterion's contribution to overall discounted performance and provide a transparent basis for ranking criteria according to their economic impact. 
The proposed approach effectively bridges theoretical discounting principles with empirical data, improving interpretability and supporting better-informed decision-making in manufacturing performance evaluation. 
\end{abstract}

\noindent\rule{16 cm}{1.3pt}


\noindent\rule{16cm}{1.3pt}

\section{Introduction}
Manufacturing firms operate in highly competitive, capital-intensive environments where accurate financial performance evaluation is essential. Traditional financial analysis methods often rely on historical ratios and linear models, which fail to capture the complex, non-linear relationships inherent in financial data. Artificial intelligence (AI), particularly machine learning (ML), provides advanced tools that enhance the accuracy, speed, and reliability of financial performance evaluation.

Artificial intelligence (AI) refers to the capability of computer systems or machines to perform tasks that typically require human intelligence~\cite{Schemmer2022}, such as learning, reasoning, problem-solving, perception, and decision-making. AI systems analyze data, recognize patterns, and generate predictions or recommendations with minimal human intervention. Artificial intelligence (AI) ranks among the most recent and fundamental developments in the convergence of electronic markets~\cite{Alt2021} and has emerged as an increasingly vital topic in information systems (IS) research~\cite{AbdelKarim2021,Alt2018}.

In economics and business, AI enhances efficiency, reduces costs, boosts productivity, and aids strategic decision-making. In manufacturing firms, it analyzes financial and operational data to assess performance, forecast outcomes, and manage risks~\cite{Mehrotra2019}.
Machine learning, a subset of artificial intelligence, enables systems to learn automatically from data and enhance performance over time without explicit programming. Rather than relying on fixed rules, ML models detect patterns and relationships in large datasets.

Artificial Intelligence (AI) and Machine Learning (ML) are closely related~\cite{Fahle2020}, with AI encompassing the broader field and ML constituting a key subset. AI involves developing systems that perform tasks typically requiring human intelligence, such as reasoning, decision-making, and problem-solving. ML equips~\cite{Kino2021} these systems to learn from data and enhance performance over time without explicit programming. By leveraging ML algorithms, AI systems identify patterns, analyze large datasets, and generate accurate predictions. In economic and financial contexts---particularly within manufacturing firms---ML provides the analytical capabilities enabling AI to assess financial performance, forecast outcomes, detect risks, and facilitate informed decision-making.

Financial performance evaluation in manufacturing firms entails the systematic assessment of how effectively a firm employs its financial and productive resources to attain profitability, efficiency, and long-term sustainability. This process involves analyzing financial statements through key indicators, including profitability ratios, liquidity ratios, efficiency ratios, solvency ratios, and cash flow measures. In manufacturing contexts, financial performance is significantly shaped by production efficiency, cost control, inventory management, capital investment, and technology adoption. Robust evaluation enables managers and stakeholders to pinpoint strengths and weaknesses, optimize resource allocation, manage costs, bolster competitiveness, and facilitate informed economic and strategic decisions.

The role of Artificial Intelligence (AI) in financial performance evaluation lies in its ability to analyze large volumes of financial and operational data quickly and accurately, providing deeper insights than traditional methods. AI techniques, particularly machine learning algorithms, can identify complex and non-linear relationships among financial indicators, forecast future profitability and cash flows, detect inefficiencies and financial risks, and uncover hidden patterns in manufacturing operations. By integrating financial data with operational variables such as production levels, inventory usage, and cost structures, AI enhances the accuracy, timeliness, and reliability of financial performance evaluation, thereby supporting better managerial decision-making and improved economic performance in manufacturing firms.

%==================================
 \section{Preliminaries}\label{sec2}
%==================================
The machine learning approach to evaluating financial performance employs data-driven algorithms to analyze and predict outcomes for manufacturing firms. It starts with collecting financial and operational data, followed by preprocessing steps like cleaning, normalization, and feature selection to ensure data quality. Models are then trained using algorithms such as regression, decision trees, random forests, or neural networks to identify patterns and relationships. These models undergo performance evaluation before application in forecasting financial indicators, detecting inefficiencies, and informing strategic decisions. By capturing complex data relationships, this method yields more accurate and reliable results than traditional techniques. Machine learning algorithms fall into three main categories:
\begin{itemize}
\item \textbf{Supervised learning}: Uses labeled data to predict outcomes (e.g., profit forecasting).
\item \textbf{Unsupervised learning}: Identifies hidden patterns or groupings (e.g., cost structure analysis).
\item \textbf{Reinforcement learning}: Learns through trial and error to optimize decisions.
\end{itemize}

The use of artificial intelligence (AI) and machine learning in evaluating financial performance carries significant economic implications, especially for manufacturing firms. AI-driven analyses enhance resource allocation efficiency by delivering accurate, timely financial insights; they reduce information asymmetry between managers and stakeholders; and they boost productivity via superior cost control and operational efficiency. By helping firms anticipate financial risks and optimize decision-making, AI fosters greater competitiveness and profitability. On a broader scale, AI adoption in manufacturing drives industrial growth, innovation, and sustainable economic development.

Dimensionality~\cite{Joshi2019} confuses when handling datasets. Physically, it means spatial dimensions: length, width, height---rarely more than three in reality. Machine learning data often has tens or hundreds of dimensions. The key property is orthogonality: dimensions perpendicular to each other ensure unique point representations. Without it, points have multiple coordinates, breaking calculations. The origin is $(0,0,0)$; $(1,1,1)$ is uniquely 1 unit away along each axis. Higher dimensions, like a fourth orthogonal to the first three, follow suit: origin $(0,0,0,0)$, prior point $(1,1,1,0)$. Orthogonality guarantees uniqueness for any number of dimensions.

\subsection{problem statement}

Despite extensive research on the financial performance of manufacturing firms, existing studies remain fragmented. Theoretical time-discounted economic models rigorously incorporate the time value of money but rely on overly simplistic assumptions that overlook multidimensional and interdependent decision criteria. In contrast, expert-based weighting methods, such as the Analytic Hierarchy Process (AHP), yield descriptive rankings but lack empirical validation of their influence on actual discounted performance outcomes. This fragmentation points to a critical gap in the literature: the absence of integrated frameworks that quantitatively bridge time-discounted economic theory with data-driven analysis of multiple evaluation criteria.

The present study addresses this gap by integrating discounted cost--benefit analysis with linear regression. It employs expert-derived Average Weight Scores (AWS) as explanatory variables to empirically estimate the marginal contribution of each criterion to overall discounted economic performance, thereby providing a transparent and interpretable approach for financial evaluation in manufacturing.

%=====================%%%%%====
\section{Linear Regression of Assessing Financial Statement Risks among MCDM Techniques}
\begin{figure}[H]
\centering
\includegraphics[width=0.85\textwidth]{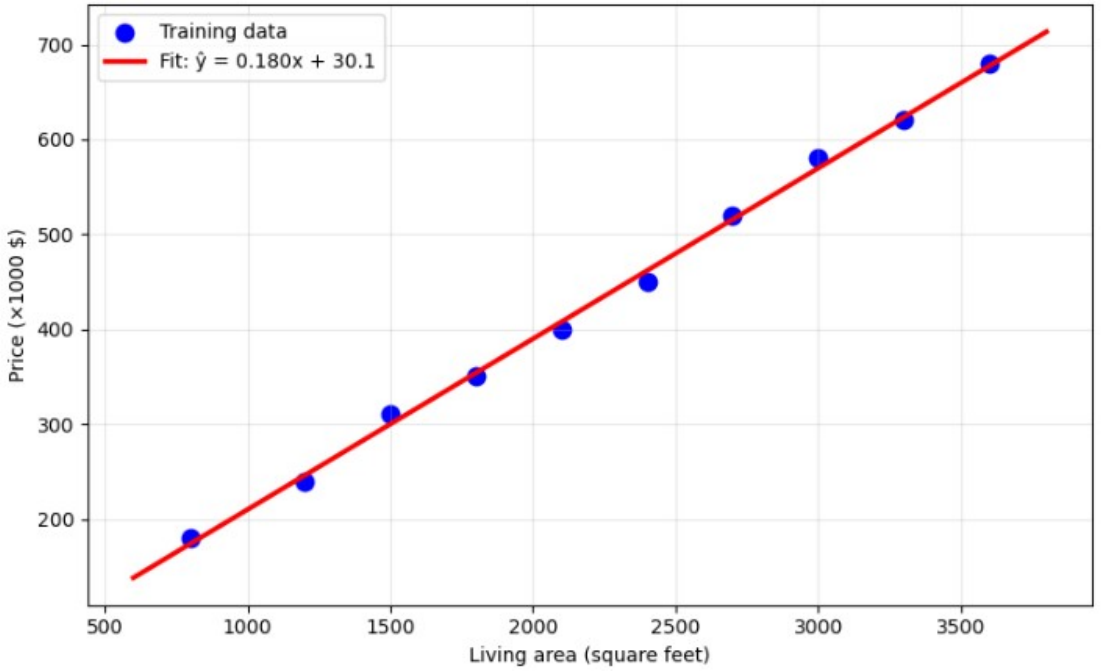}
\caption{Linear regression example among Table~\ref{tabeqq1}.}
\label{fig001linregexample}
\end{figure}

%=====================%%%%%====
Linear regression is a machine learning method that is used to model the relationship between one or more independent variables and a target variable. However, this method was not originally designed to solve classification problems and may have limitations in
situations with class imbalance, as it happens in medical problems where it is important
to correctly identify rare cases of the disease~\cite{Khizhnyak2015,Alekseeva2024}.
Linear regression exemplifies strictly linear models and is also known as polynomial fitting~\cite{Joshi2019}, representing one of the simplest methods in machine learning. Consider a linear regression problem with training data comprising $p$ samples, where the inputs are $n$-dimensional vectors $(x_i)_{i=1,\ldots,p}$ with each $x_i \in \mathbb{R}^n$, and the outputs are scalars $(y_i)_{i=1,\ldots,p}$ with each $y_i \in \mathbb{R}$. The method of linear regression defines the relationship between inputs $x_i$ and predicted outputs $\hat{y}_i$ via the linear equation
\begin{equation}~\label{eqq1kin}
\hat{y}_i = \sum_{j=1}^{n} x_{ij} w_j + w_0.
\end{equation}
Here, $\hat{y}_i$ denotes the predicted output corresponding to the actual output $y_i$. The parameters $w_j$ ($j = 1, \dots, n$) are the weights, and $w_0$ is the bias. Training aims to determine these parameters. Equivalently, in matrix form,
\begin{equation}~\label{eqq2kin}
\hat{y} = X^T w + w_0,
\end{equation}
where $X = [x_1, \dots, x_p]^T$ and $w = [w_1, \dots, w_n]^T$. The goal is to estimate these parameters using the training data. The estimator derived in this manner is known as the \textit{maximum likelihood estimator (MLE)}, which is the optimal unbiased estimator given the training data. The corresponding optimization problem is
\begin{equation}~\label{eqq3kin}
\min_{\mathbf{w}} \sum_{i=1}^n (y_i - \hat{y}_i)^2
\end{equation}
where $\hat{y}_i$ denotes the predicted value. Substituting $\hat{y}_i = \mathbf{w}^\top \mathbf{x}_i + w_0$, the minimization problem for the optimal weight vector $\mathbf{w}^*$ becomes
\begin{equation}~\label{eqq4kin}
\mathbf{w}^* = \arg\min_{\mathbf{w}, w_0} \sum_{i=1}^n \Bigl( y_i - \Bigl( \sum_{j=1}^p w_j x_{ij} + w_0 \Bigr) \Bigr)^2.
\end{equation}
This constitutes a standard quadratic optimization problem, extensively studied in the literature. As the formulation relies entirely on linear equations, it can model only linear relationships between inputs and outputs. To illustrate the concepts introduced above, consider a simple real-world application of linear regression: predicting house prices from living area. Suppose we have a small training dataset comprising $p=10$ houses. Each house is described by a single input feature ($n=1$): its living area $x_i$ in square feet. The corresponding target variable $y_i$ represents the house price in thousands of dollars among Table~\ref{tabeqq1}. A subset of the data might appear as follows:
\begin{table}[H]
\centering
\begin{tabular}{c|c|c||c|c|c}
\hline
House $i$ & Size $x_i$ (sq ft) & Price $y_i$ ($\times 1000\ $) & House $i$ & Size $x_i$ (sq ft) & Price $y_i$ ($\times 1000\ $) \\
\hline
1  &  800  & 180 & 2  & 1200  & 240 \\ \hline
3  & 1500  & 310& 4  & 1800  & 350 \\ \hline
5  & 2100  & 400 & 6  & 2400  & 450 \\ \hline
7  & 2700  & 520 & 8  & 3000  & 580 \\ \hline
9  & 3300  & 620 & 10 & 3600  & 680 \\ \hline
\end{tabular}
\caption{Example training data: house size vs.\ price}~\label{tabeqq1}
\end{table}
Linear regression assumes a strictly linear relationship between the input feature and the output variable. Accordingly, we model the predicted house price $\hat{y}_i$ as
\begin{equation}\label{eq01housemodel}
\hat{y}_i = w_1 x_i + w_0, 
\end{equation}
where $w_1$ represents the slope (weight) and $w_0$ is the intercept (bias term). The training objective is to find the parameters $w_1$ and $w_0$ that minimize the sum of squared errors over the training set:
\begin{equation} \label{eq02houseoptimization}
\min_{w_1, w_0} \sum_{i=1}^{10} (y_i - \hat{y}_i)^2.
\end{equation}
This formulation is precisely the ordinary least-squares problem. Under the assumption of additive Gaussian noise, the resulting parameter estimates constitute the maximum likelihood estimator (MLE). A typical fitted model for this dataset might be approximately
$$
\hat{y} \approx 0.18 \cdot x + 40 \quad \text{(i.e., } w_1 \approx 0.18,\ w_0 \approx 40\text{)}.
$$
This implies that, on average, each additional square foot of living area is associated with an increase of approximately \$180 in house price (since $y_i$ is measured in thousands of dollars). Figure~\ref{fig001linregexample} displays the training data points together with the fitted regression line. The line represents the linear model that minimizes the sum of squared vertical distances between the observed points and the predictions.

Note that, blue points show the training data (house size versus price). The red line is the fitted model $\hat{y}_i = w_1 x_i + w_0$ obtained by minimizing the sum of squared errors, which corresponds to the maximum likelihood estimator under Gaussian noise. Because the model is strictly linear, it cannot capture nonlinear patterns, such as diminishing returns at very large house sizes. This example illustrates why linear regression remains one of the simplest and most interpretable methods in machine learning, while also highlighting its fundamental limitation: it can only model strictly linear relationships between inputs and outputs.

We now examine methods for calculating control system efficiency, incorporating the time factor: frozen funds harm the national economy, while early commissioning yields benefits. The timing of benefits (or costs) matters---the later the effect, the lower its present value. For instance, one ruble saved in ten years equals 39 kopecks today, and one ruble saved in twenty years equals 15 kopecks today. To address this, we discount effects at different times to a single moment using the benefit function $\theta(t)$. The discounted average benefit over period $T$ is
\begin{equation}~\label{eqq5kin}
\bar{R}=\frac{1}{T}\left[\Sigma R(\theta) t+\int_{0}^{T} r \theta(t) d t\right]
\end{equation}

where $R$ is the one-time income (expense) at time $t$, and $r$ is the current income (expense) per unit time at time $t$. In the simple case from~\eqref{eqq5kin}, the benefit function follows the compound interest formula:
\begin{equation}~\label{eqq6kin}
\theta(t)=(1+\alpha)^{-t}
\end{equation}

Revenues (or costs) may be: 1) one-time, 2) proportional to time, or 3) exhibit complex time dependence.
In the first case, revenues (costs) according to~\eqref{eqq6kin} brought to $t=0$ equal $\Sigma R(\theta) t$.
The second case, such as operating costs, is characterized by annual savings of $I$ rub/year. The total savings over time $t$, discounted to the startup moment, is
\begin{equation}~\label{eqq7kin}
r^{\prime}=\int_{0}^{t} I(1+\alpha)^{-t} \, dt
\end{equation}
According to~\eqref{eqq7kin}, by denoting $\tau=1 / \ln (1+\alpha)$, yields
\begin{equation}~\label{eqq8kin}
r^{\prime}=I \tau\left(1-e^{-t / \tau}\right).
\end{equation}
For $\alpha \leqslant 0.15$, $\tau=1.02 / \alpha$ (error within $\pm 3\%$).
The third case is illustrated by costs to create a control system. Figure 1 plots costs (from reporting data) for ten systems, from facility survey to startup, spanning 1.5--6 years. The bold line shows the averaged curve. For simplification, total costs can be attributed to the ``center of gravity'' of yearly costs. For the averaged curve in Fig. 1, accounting for compound interest, this center is approximately $0.5 t_{\mathrm{y}}$, where $t_{\mathrm{y}}$ is the creation and startup duration (in years). Thus, the cost at startup $C_{\mathrm{y}}^{\prime}$, according to~\eqref{eqq8kin}
\begin{equation}~\label{eqq9kin}
C_{\mathrm{y}}^{\prime}=C_{\mathrm{y}}(1+\alpha)^{0.5 t_{\mathrm{y}}},
\end{equation}
where $C_{\mathrm{y}}$ is the total cost (without interest).

Ridge regression~\cite{Joshi2019} augments the ordinary least-squares objective with an $\ell_2$ penalty on the coefficients from~\eqref{eqq4kin}. Its constrained formulation is
\begin{equation}~\label{eqq10kin}
\min_{\mathbf{w}, w_0} \sum_{i=1}^n (y_i - \mathbf{x}_i^\top \mathbf{w} - w_0)^2
\quad \text{subject to} \quad \|\mathbf{w}\|_2^2 \leq t,
\end{equation}
where $t > 0$ is the constraint parameter and $n$ is the number of observations. 
Through the Lagrangian, this constrained problem is equivalent to the following unconstrained optimization:
\begin{equation}~\label{eqq11kin}
(\mathbf{w}_{\text{Ridge}}, w_0) = \arg\min_{\mathbf{w}, w_0} \left\{
\sum_{i=1}^n (y_i - \mathbf{x}_i^\top \mathbf{w} - w_0)^2 + \lambda \|\mathbf{w}\|_2^2
\right\},
\end{equation}
where $\lambda \geq 0$ is the regularization parameter (Lagrange multiplier). In matrix notation, the objective becomes
$$
\min_{\mathbf{w}, w_0} \ \| \mathbf{y} - \mathbf{X}\mathbf{w} - w_0 \mathbf{1}_n \|_2^2 + \lambda \|\mathbf{w}\|_2^2.
$$

Thus, from~\eqref{eqq10kin} and \eqref{eqq11kin}, the ridge penalty shrinks the coefficient estimates toward zero---though it does not set them exactly to zero---thereby improving numerical stability and generalization performance, particularly when features are highly correlated or when the number of observations $n$ is not substantially larger than the number of predictors $p$.

\begin{figure}[H]
    \centering
    \includegraphics[width=0.7\linewidth]{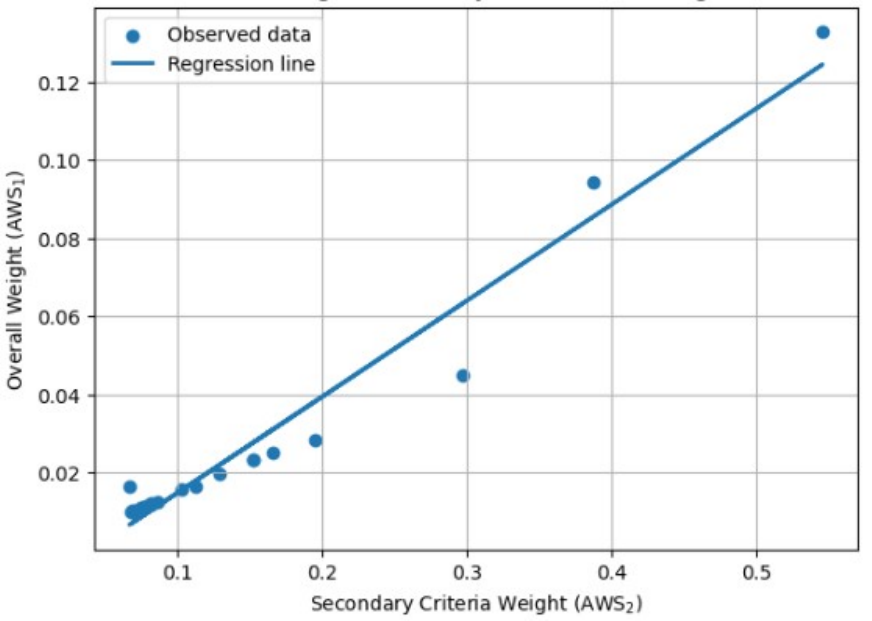}
    \caption{Linear regression analysis between secondary criteria weights ($AWS_2$) and overall weights ($AWS_1$)}
    \label{fig002linregexample}
\end{figure}

Furthermore, according to Abdullah, M. et al.~\cite{Abdullah2025}, let the dependent variable be the overall importance score of each criterion, denoted by $AWS_1$. The independent variables are the secondary-criteria weight ($AWS_2$) and the main-criteria weight ($AWS_3$). For each observation $i = 1, 2, \ldots, p$, the linear regression model is defined as:
\begin{equation}~\label{eqq12kin}
AWS_{1,i} = \beta_0 + \beta_1 AWS_{2,i} + \beta_2 AWS_{3,i} + \varepsilon_i
\end{equation}
where $\beta_0$ represents the intercept, $\beta_1$ and $\beta_2$ are the regression coefficients, and $\varepsilon_i$ is the random error term. The linear regression model can be expressed in matrix form as:
\begin{equation}~\label{eqq13kin}
\mathcal{Y} =\mathcal{X}\beta + \varepsilon
\end{equation}
where
\begin{equation}
\mathcal{X} =
\begin{bmatrix}
1 & AWS_{2,1} & AWS_{3,1} \\
1 & AWS_{2,2} & AWS_{3,2} \\
\vdots & \vdots & \vdots \\
1 & AWS_{2,p} & AWS_{3,p}
\end{bmatrix},
\quad
\beta=
\begin{bmatrix}
\beta_0 \\
\beta_1 \\
\beta_2
\end{bmatrix}
\end{equation}
and $\mathcal{Y} = [AWS_{1,1}, AWS_{1,2}, \ldots, AWS_{1,p}]^\top$. The regression coefficients are estimated using Ordinary Least Squares (OLS), which minimizes the sum of squared residuals between the observed and predicted values. The OLS estimator is given by:

$$
\hat{\beta} = \arg\min_{\beta} \sum_{i=1}^n (y_i - x_i'\beta)^2
$$
Thus, we obtain among Table~\ref{tab002top} the following results
\begin{table}[H]
\centering
\begin{tabular}{c S[table-format=1.4] S[table-format=1.4] S[table-format=1.4] l}
\hline
{Global Rank} & {Actual weight} & {Predicted} & {Difference} & {Comment} \\
\hline
1 & 0.1329 & 0.0591 & +0.0738 & \textcolor{red!70!black}{\textbf{heavily underpredicted}} \\
2 & 0.0944 & 0.0573 & +0.0371 & significantly underpredicted \\ \hline
3 & 0.0450 & 0.0555 & -0.0105 & slightly overpredicted \\ \hline
4 & 0.0426 & 0.0537 & -0.0111 & slightly overpredicted \\ \hline
\hline
\end{tabular}
\caption{Actual vs. predicted weights for top-ranked criteria}
\label{tab002top}
\end{table}
Thus, a simple linear model assumes that weights decrease \textit{nearly linearly} (almost constantly) as rank worsens.
In practice, the \textit{top 1--2 positions} (in particular rank 1 = LR3 and rank 2 = LR2) carry \textit{disproportionately large weights}.
From approximately rank~5 onward, the relationship becomes much closer to linear.
The overall $R^2 \approx 0.55$ indicates a moderate linear fit.
The coefficient $\beta_1$ quantifies the contribution of the secondary-criteria weights to the overall importance score, while $\beta_2$ captures the contribution of the main-criteria weights. Statistically significant coefficients indicate criteria that exert a substantial influence on financial performance evaluation (see Figure~\ref{fig002linregexample}).

This linear regression model serves as an interpretable baseline for understanding relationships within the hierarchical criteria structure prior to the application of more advanced machine learning methods.  The analytical framework outlined above demonstrates the economic significance of accounting for the time value of costs and benefits when assessing control system efficiency according to ~\eqref{eqq5kin}--\eqref{eqq9kin}. In practical applications, however, the discounted economic effect depends on multiple interdependent criteria whose functional relationships are not known beforehand. Linear regression is therefore employed here as an approximation method to empirically quantify these relationships.

In this study, the discounted economic performance is modeled as a linear function of the primary and secondary evaluation criteria obtained from expert assessments. The average weight scores (AWS) reported in Table [9,\cite{Abdullah2025}] serve as observable explanatory variables that reflect the relative economic importance of each criterion. The regression model is specified from~\eqref{eqq11kin}, \eqref{eqq12kin} and \eqref{eqq14kin} as
\begin{equation}~\label{eqq14kin}
\bar{R}_i = \beta_0 + \beta_1 \mathrm{CSR}_i + \beta_2 \mathrm{LR}_i + \beta_3 \mathrm{IR}_i + \beta_4 \mathrm{CFR}_i + \varepsilon_i,
\end{equation}

where $\bar{R}_i$ represents the discounted economic effect for observation $i$, $\mathrm{CSR}_i$, $\mathrm{LR}_i$, $\mathrm{IR}_i$, and $\mathrm{CFR}_i$ denote the aggregated weight scores of the respective criteria, $\beta_j$ are the regression coefficients, and $\varepsilon_i$ is the random error term. This approach allows an empirical assessment of how time-discounted economic efficiency responds to changes in the relative importance of the evaluation criteria, thereby bridging the theoretical discounting framework with expert-derived empirical data.

\section{Conclusion}

This study demonstrates the value of an integrated framework that combines theoretical time-discounted economic analysis with empirical linear regression modeling to assess the financial performance of manufacturing firms. By expressing costs and benefits in present-value terms using compound interest-based discounting functions---and then regressing the resulting discounted performance outcomes on expert-derived Average Weight Scores (AWS)---the proposed approach quantifies the marginal economic contributions of diverse criteria, including cost structure, investment risk, reliability, and cash-flow dynamics.

The results yield a transparent, data-driven ranking of evaluation criteria according to their actual influence on overall discounted performance. This overcomes key limitations of purely theoretical models, which often oversimplify complex multidimensional interactions, and of standalone expert-weighting methods, which typically lack empirical validation. The hybrid methodology improves interpretability and supports more robust, evidence-based decision-making in the evaluation of manufacturing systems and the assessment of control system efficiency.
Although the framework represents a practical advance in bridging discounting principles with regression-based insights, future research could enhance its applicability by incorporating nonlinear relationships, dynamic panel data models, or industry-specific calibrations of AWS to improve predictive accuracy and generalizability. Ultimately, this approach contributes to more comprehensive financial performance evaluation practices, enabling manufacturing firms to better align operational and managerial decisions with long-term economic value creation.

%===========================
\section*{Declarations}
\begin{itemize}
	\item Funding: Not Funding.
	\item Conflict of interest/Competing interests: The author declare that there are no conflicts of interest or competing interests related to this study.
	\item Ethics approval and consent to participate: The author contributed equally to this work.
	\item Data availability statement: All data is included within the manuscript.
\end{itemize}

\end{document}